\documentclass{jfm}

\usepackage{caption}
\usepackage{subcaption}
\usepackage{graphicx}
\usepackage{epstopdf, epsfig}
\usepackage{amsmath,bm}
\usepackage{cleveref}
\usepackage{xcolor}
\usepackage{url}
\usepackage[super]{nth}
\graphicspath{ {images/} }

\title{Vortex Dynamics: A Variational Approach Using the Principle of Least Action}

\author{Nabil M. Khalifa
	\and Haithem E. Taha\corresp{\email{hetaha@uci.edu}}}

\affiliation{Mechanical and Aerospace Engineering, University of California, Irvine,
	CA 92697, USA}

\begin{document}
\maketitle

\begin{abstract}
The study of vortex dynamics using a variational formulation has an extensive history and a rich literature. The standard Hamiltonian function that describes the dynamics of interacting point vortices of constant strength is the Kirchhoff-Routh (KR) function. This function was not obtained from basic definitions of classical mechanics (i.e., in terms of kinetic and potential energies), but it was rather devised to match the already known differential equations of motion for constant-strength point vortices given by the Bio-Savart law. Instead, we develop a new variational formulation for vortex dynamics based on the principle of least action. As an application, we consider a system of non-deforming, free vortex patches of constant-strength. Interestingly, the obtained equations of motion are second-order differential equations defining vortex accelerations, not velocities. In the special case of constant-strength point vortices, the new formulation reduces to the motion determined from the KR function. However, for finite-core vortices, the resulting dynamics are more complex than those obtained from the KR formulation. For example, a pair of equal-strength, counter-rotating vortices could be initialized with different velocities, resulting in interesting patterns which cannot be handled by the KR approach. Also, the developed model easily admits external body forces (e.g., electromagnetic). The interaction between the electrodynamic force and the hydrodynamic vortex force leads to a rich, counter-intuitive behavior that could not be handled by the KR formulation. Finally, this new variational formulation, which is derived from first principles in mechanics, can be easily extended to deforming and/or arbitrary time-varying vortices.

\end{abstract}


\section{Introduction}\label{sec:Introduction}
Vortex dynamics represents one of the main pillars of fluid mechanics. The theoretical edifice has been growing since the seminal papers of the founder \cite{helmholtz1858integrale,helmholtz1867lxiii}. He introduced the vorticity equations, and established the circulation conservation laws when the fluid motion emerges from a potential force. The analogy between potential flow and electricity, discovered by \cite{maxwell1869remarks}, has lead to simple models of vortex dynamics \citep{lamb,milneaero}. However, the potential-flow formulation is a kinematic representation of the flow field; it lacks dynamical features \citep{gonzalez2022variational}. For example, the standard law  that is ubiquitously used in the literature to describe the motion of two-dimensional vortices (infinite vortex filament) is the \textit{Biot-Savart law}, which provides first-order ODEs in the positions of the vortices, lacking any dynamical representation that considers forces and provides accelerations accordingly; no initial velocity of the vortices can be assigned, in contrast to any typical dynamics problem. A true fluid \textit{dynamic} model would result in second-order ODEs in the position of the vortices.
  
\cite{kirchhoff1883vorlesungen} studied free N-point vortices of constant strength in an unbounded domain, and described their motion based on the Biot-Savart law. Interestingly, the resulting ODEs possessed a Hamiltonian structure \citep{goldstein2002classical,lanczos2020variational}. Based on this finding, they defined a Hamiltonian function (the Kirchhoff function) for constant-strength point vortices, which is the standard Hamiltonian of vortex motion in the literature \citep{aref2007playground}. However, this Hamiltonian was not derived from basic definitions or first principles; it is not defined in terms of the physical quantities (kinetic and potential energies) that typically constitute the Hamiltonian function for a mechanical system. It was simply constructed from the reciprocal property of the stream function of point vortices. Therefore, its use outside this case is questionable, and its extension to other scenarios may not be clear except for a few cases as described below.

To account for the interaction with a body, \cite{routh1880some} extended Kirchhoff's model using the method of images. Afterwards, \cite{cclinI,cclinII} provided the most generic and complete formulation of point vortices interacting with a body in two dimensions. He was able to show the existence of a hybrid function $W$ composed of Kirchhoff's and Routh's, which described the Hamiltonian dynamics of a system of vortices and called it the \textit{Kirchhoff-Routh} function. Arguably, the $W$ function could be considered as the underlying foundation of every application and study that includes constant-strength point vortices \citep{vortex_methods,theory_of_concentrated_vortices,aref2007playground}. It is interesting to mention that  \cite{cclinI,cclinII} relied on the concept of  \textit{generalized Green's} function, which was extended later by \cite{daren_crowdy} to multiply-connected domains relying on prime functions (see Section 16 in \citep{daren_crowdy} for the extended KR function to multiply-connected domains). Nevertheless, in all these representations \citep{kirchhoff1883vorlesungen,routh1880some,cclinI,cclinII}, the resulting ODEs were first-order in the position variables, lacking the dynamic essence of mechanical systems which are typically described by second-order ODEs defining accelerations.

Though peripheral to the main focus of this paper, it is perhaps worthy to mention the classical work of \cite{grobli1877specielle} on the three- and four-point vortices problem. Interestingly, using the kinematic formulation of the motion of point vortices enabled by the Biot-Savart law, resulting in first-order ODEs in position, he showed that the three-vortex problem is integrable! This problem was studied further by several authors such as \cite{synge1949motion,rott1989three,rott1990constrained,aref1979motion,aref1989three}.  Moreover, \cite{grobli1877specielle} was able to solve the integrable case of four vortices equal in magnitude and arranged as a cross-section of two coaxial vortex rings, in which he was able to predict the leapfrogging phenomenon.  However, the four-vortex case is not always integrable and nonlinearities may complicate the dynamics,  introducing chaotic behavior, as studied extensively by  \cite{aref1984stirring,aref1983integrableannualreview,aref1980integrable4vortices,aref1982integrable4vorticesI,arefeckhardt1988integrable4vorticesII,aref1999fourzerocirc}. This discussion presents the importance of the three-vortex problem as a classical problem, especially since \cite{newton2013n} showed that the behavior of any N-vortices could be studied as a multiple of vortex triads, as shown in his detailed monograph describing most of the analytical studies of point vortices \citep{newton2013n}. Despite of all numerous contribution to the three- and four- point vortices problem described previously, the work of Gr{\"o}bli still remains a classic contribution \citep{aref1992grobli}. 

The kinematic formulation based on the Biot-Savart law has been extensively used in the literature for numerical simulation of vortex patches by discretizing them into many point vortices \citep{vortex_methods,theory_of_concentrated_vortices}. However, this approach has its drawbacks because the resulting kernel for integration is singular, which requires regularization. The \textit{Vortex Blob} \citep{leonard1980vortex} method was introduced to relax the highly singular nature of point vortices and define smoother velocity fields. The method relied on defining a vortex core shape with a vorticity distribution inside it. Moreover, the vortex blob geometry was held constant throughout the motion --- an assumption that could introduce spatial errors. However, \cite{leonard1980vortex} obtained an estimate of such an error while \cite{hald1978convergence} and \cite{hald1979convergence} showed that the resulting motion converges to Euler's solution. The approach of \textit{Contour Dynamics} was proposed as an alternative that does not suffer from singularity issues and spatial errors. It relies on the fact that vortex patches are solutions of Euler's equations, hence, the boundary of the vortex patch is considered to be a material derivative. As a consequence, one needs only to track the velocity induced by the other vortices over this boundary and, hence, update the vortex position accordingly. The contour dynamics technique was first introduced by \cite{zabusky1979contour} and multiple efforts were subsequently exerted by \cite{saffman1980equilibrium_shapes_pair_vortex_patches,pullin1992contour_anual_review} and \cite{saffman}. The main concept behind the contour dynamics approach is solely a kinematical one --- again  highlighting that this approach inherently lacks  kinetics.

The study of point vortices interacting with a body has enjoyed significant interest due to its relevance to many applications; e.g., the unsteady evolution of lift over an airfoil due to the interaction of the \textit{starting vortex} with the body \citep{tchieu2011discrete,taha2020high}. \cite{ramodanov2001onevortexandcylinder} studied the interaction of a point vortex with a cylinder (the F{\"o}ppl problem), then \cite{borisov2004integrability} inspected the integrability of this problem. Moreover, \cite{borisov2005RC,borisov2005dynamics,borisov2007JMF} and \cite{ramodanov2021dynamics} extended this setup to allow for multiple point vortices to interact with a cylinder. In another classic article by \cite{shashikanth2002hamiltonian}, the authors investigated the Hamiltonian dynamics of a cylinder interacting with a number of point vortices. In all of these efforts, the $W$ function was considered as the main underlying foundation --- the Hamiltonian function for the motion of constant-strength point vortices. However, because the $W$ function was not derived from the basic definitions or first principles in mechanics, this approach collapses if time-varying point vortices are considered. The KR dynamics will simply fail; while a point vortex of constant-strength moves with the local flow velocity (the Kirchhoff velocity) according to Helmholtz laws of vortex dynamics \citep{helmholtz1858integrale}, a vortex with time-varying strength does  not, as it is well known in the unsteady aerodynamics literature \citep{eldredge,wang2013low,michelin2009unsteady,tchieu2011discrete,tallapragada2013reduced}. This dilemma was resolved by \cite{brown1954effect}, who developed a convection model (also first-order in the position variables) for time-varying point vortices. An alternative model, named the \textit{Impulse Matching} model, was proposed by \cite{tchieu2011discrete} and \cite{wang2013low}. An intuitive reasoning for the failure of the  KR function in predicting the motion of unsteady point vortices, is that, formally, it does not contain any information beyond the Biot-Savart law, which describes flow kinematics without any kinetics. This argument reveals one of the needs for a true dynamical framework capable of describing vortex dynamics from first principles, which is the main goal of this article.

While there have been numerous efforts on the Hamiltonian formulation of point vortices, little has been done to develop a Lagrangian framework. This fact may be intuitive given the lack of veritable dynamics in these Hamiltonian formulations; a true mechanical system will possess both Lagrangian and Hamiltonian formulations, mutually related through the Legendre transformation \citep{goldstein2002classical}. However, having an arbitrary Hamiltonian function that happens to reproduce a given set of ODEs does not necessarily represent a mechanical system. As \cite{salmon1988hamiltonian} put it, ``\textit{the existence of a Hamiltonian structure is, by itself, meaningless because any set of evolution equations can be written in canonical form}". These non-standard Hamiltonians may not be associated with Lagrangian functions. However, there are a few exceptions. For example, \cite{chapmanlagrangian} devised a Lagrangian function that reproduces the kinematic set of ODEs dictated by the KR Hamiltonian. However, this Lagrangian was not derived from basic definitions or first principles in mechanics (i.e., it is not defined based on the kinetic and potential energies of the system), so it had to be in a non-standard form (e.g., bi-linear in velocity, not quadratic) to result in first-order ODEs, in contrast to the second-order ODEs that typically result from a standard Lagrangian. In a similar spirit, \cite{hussein2018variational} introduced a Lagrangian that reproduces the first-order ODEs of the Brown-Michael model, describing the motion of point vortices with time-varying strength; they used it to study the motion of the starting vortex behind an airfoil and its effect on the lift evolution similar to \cite{wang2013low} and \cite{tchieu2011discrete}.

Perhaps the only study of vortex motion that showed a mechanical structure (i.e., second-order ODEs in position) was done by \cite{ragazzodensevortex_electro_simmilarity}, who considered point vortices with finite masses. Using an analogy with electromagnetism, they constructed a Hamiltonian of the dense vortices based on that of point charges in a magnetic field. Interestingly, the non-zero inertia of the point vortices (similar to the non-zero charge) resulted in second-order ODEs in the position variables, leading to a richer behavior than that obtained using the KR kinematics. However, the fact that the authors relied completely on analogy between hydrodynamics and electrodynamics to develop their dynamic model, perhaps sets no difference between it and the devised KR Hamiltonian as far as first principles are concerned. On a promising side, \cite{olva1996motion_ragazzo_perturbation} was able to show that their resulting ODEs reduce to those of the KR kinematics in the limit to a vanishing vortex mass, which, in turns, shows that the KR kinematics is a singular perturbation of \cite{ragazzodensevortex_electro_simmilarity} dynamics.

While we did our best to present the relevant literature to the reader, it is certain that we are far from providing a complete account of the perhaps unfathomable literature on the topic. Nor is it our goal here to provide a comprehensive review of such a rich topic. For a thorough review and more information about point vortex dynamics, the reader is referred to multiple sources \citep{meleshko2007bibliography,aref2007playground,smith2011singularities,newton2014post_aref_era,lamb,batchelor,saffman,milnehydro,milneaero,eldredge,wu_red_book_vorticity,theory_of_concentrated_vortices,vortex_methods,truesdell2018kinematics}.

In conclusion, although the previous efforts, discussed above, were quite legitimate and spawned very interesting results \citep[see][]{aref2007playground}, they suffer from the following drawbacks: (1) the defined Hamiltonian and Lagrangian functions were neither derived from basic definitions in mechanics nor from the available rich literature on variational principles in fluid mechanics \citep{seliger1968variational,bateman1929notes,salmon1988hamiltonian,herivel1955derivation,bretherton1970note,penfield1966hamilton,hargreaves1908xxxvii,serrin1959mathematical,luke1967variational,loffredo1989eulerian,morrison1998hamiltonian,berdichevsky_book}; (2) They were either devised to reproduce the already known ODEs or developed based on an analogy with another discipline; (3) As such, the resulting ODEs are merely the kinematic ones of the Biot-Savart law; no rich dynamics can be obtained; (4) The fact that these functions were not derived from first principles does not allow extension to cases beyond free, constant-strength, point vortices.  Indeed, one would hope to see a straightforward derivation of the Hamiltonian (or Lagrangian) of point vortices from basic definitions of mechanics (i.e., from kinetic and potential energies) or from the rich legacy of variational principles of continuum fluid mechanics \citep{seliger1968variational,bateman1929notes,salmon1988hamiltonian,herivel1955derivation,bretherton1970note,penfield1966hamilton,hargreaves1908xxxvii,serrin1959mathematical,luke1967variational,loffredo1989eulerian,morrison1998hamiltonian,berdichevsky_book}, which is the main goal of this work. Such a model will resolve the drawbacks listed above, resulting in second-order dynamics, allowing extension to unsteady vortices, and admitting arbitrary forces (e.g., gravity, electromagnetic, etc).

In this paper, we rely on the principle of least action and its application to the dynamics of a continuum inviscid fluid by \cite{seliger1968variational}. Exploiting this variational principle, we develop a novel formulation for vortex dynamics of circular patches (Rankine vortices). In contrast to previous efforts, the defined Lagrangian and the dynamical analysis are derived from first principles. Interestingly, the new formulation results in a second-order ODE defining vortex accelerations, consequently allowing for richer dynamics if the initial velocity is different from the Biot-Savart induced velocity. Moreover, the resulting dynamics recover the KR kinematic ODEs in the limit of a vanishingly small core. Also, since the model emerges from a formal dynamical analysis, it could take into account different body forces arising from a potential energy. Such a capability is demonstrated in this study by considering an electromagnetic force acting on charged vortices cores. The resulting interaction between the electrodynamic force and the hydrodynamic vortex force leads to a rich, nonlinear behavior that could not be captured by the standard KR formulation. Finally, this new variational formulation can easily be extended to arbitrary time-varying vortices with deforming boundaries. However, for brevity and clarity, these extensions will be considered in future work.

\section{Hamiltonian of Point Vortices: the Kirchhoff-Routh Function}
In this section, the KR function is presented, demonstrating its Hamiltonian role for constant-strength point vortices. The KR function for free, constant-strength point vortices in an unbounded flow is given by \citep{batchelor}
\begin{equation}\label{eq:KR}
	W=\frac{-1}{4\pi}\underset{i\neq j}{\sum_{i}\sum_{j}}\,\Gamma_i\Gamma_j\log r_{ij},
\end{equation}
where $\Gamma$'s represent the strengths of the vortices, and $r_{ij}$ is the relative distance between the $i^{\rm{th}}$ and $j^{\rm{th}}$ vortices. To show how this function serves as the Hamiltonian of free, constant-strength point vortices, let us recall the stream function describing the flow field at any particular point $(x,y)$ in the domain
\begin{equation}\label{eq:Stream function}
	\Psi(x,y)=-\frac{1}{2\pi}\sum_{i}\Gamma_i \log r_i,
\end{equation}
where $r_i$ is the distance between the point $(x,y)$ and the $i^{\rm{th}}$  vortex. As such, the total induced velocity at the $j^{\rm{th}}$ vortex (ignoring the vortex self-induction) is given by
\begin{equation}\label{eq:Induced_velocity}
	u_j=\frac{dx_j}{dt}=\frac{-1}{2\pi}\sum_{(i\neq j)}\Gamma_i\frac{(y_j-y_i)}{r_{ij}^2},\qquad v_j=\frac{dy_j}{dt}=\frac{1}{2\pi}\sum_{(i\neq j)}\Gamma_i\frac{(x_j-x_i)}{r_{ij}^2}.
\end{equation}
Then, multiplying \cref{eq:Induced_velocity} by $\Gamma_j$ will allow writing its right hand side in terms of the scalar function $W$ as
\begin{equation}\label{eq:F_before_trans}
	\Gamma_j\frac{dx_j}{dt}=\frac{\partial W}{\partial y_j},\qquad \Gamma_j\frac{dy_j}{dt}=\frac{-\partial W}{\partial x_j}.
\end{equation}
Defining $q_j=\sqrt{\Gamma_j}x_j$ and $p_j=\sqrt{\Gamma_j}y_j$, it is clearly seen that if $\Gamma_j$ is constant, then the ODEs (\ref{eq:F_before_trans}) is in the Hamiltonian form
\begin{equation}\label{eq:KR_as_Hamiltonian}
	\dot{q}_j=\frac{\partial W}{\partial p_j},\qquad
	\dot{p}_j=\frac{-\partial W}{\partial q_j},
\end{equation}
with $W$ serving as the Hamiltonian.

As demonstrated above, the KR Hamiltonian is not derived from basic definitions of mechanics in terms of kinetic and potential energies. This non-standard Hamiltonian merely reproduces the Biot-Savart kinematic equations in the Hamiltonian form (\ref{eq:KR_as_Hamiltonian}). However, one can relate this function $W$ to the regularized Kinetic Energy (KE) as shown in \citep{batchelor,saffman,lamb,milnehydro}. The KE ($T$) of the flow field is given by
\begin{equation}\label{eq:regularized_ke}
	T=-\frac{\rho}{4\pi}\underset{i\neq j}{\sum_{i}\sum_{j}}\,\Gamma_i\Gamma_j\ln r_{ij}-\frac{\rho}{4\pi}\left(\sum\Gamma_i^2\right)\ln \epsilon+\frac{\rho}{4\pi}\left(\sum\Gamma_i\right)^2\ln R, \qquad R\rightarrow\infty,\epsilon\rightarrow 0,
\end{equation}
where $\epsilon$ is a small radius of a circle centered at every point vortex and $R$ is the radius of the external boundary, which  extends to infinity. It is clearly seen that the first term is exactly the $W$ function scaled by the density $\rho$ and the last two terms are unbounded as $R\rightarrow \infty$ and $\epsilon \rightarrow 0$. However, these two unbounded terms are constants and do not depend on the co-ordinates (i.e., positions of the vortices). As a result, the first term (the \textit{regularized} KE) represents the only variable portion of the infinite KE and was satisfactorily considered as the Hamiltonian of point vortices, while the last two terms are dropped. The previous analysis is notably valid only for constant-strength point vortices and suffers from the several drawbacks discussed in the previous section. Alternatively, we rigorously develop a novel model for vortex dynamics from formal variational principles of continuum fluid mechanics, developed by \cite{seliger1968variational}.

\section{Variational Approach and the Principle of Least Action}
\subsection{The Principle of Least Action}
Variational principles have not been popular for the past decades in the fluid mechanics literature. \cite{penfield1966hamilton} and \cite{salmon1988hamiltonian} asked ``\textit{Why Hamilton's principle is not more widely used in the field of fluid mechanics}?” The main reason---from our perspective---is that most of these principles are based on the principle of least action, which is essentially applicable to particle mechanics. So, it can be easily extended to continuum fluid mechanics in the Lagrangian formulation. However, extension to the convenient Eulerian formulation invokes the introduction of artificial variables (e.g., Clebsch representation of the velocity field,  \cite{clebsch1859ueber}) and imposing additional constraints (Lin’s constraints, \cite{lin1961hydrodynamics}). So, in many times, these variational formulations in the literature are imbued with a sense of ad-hoc and contrived treatments, which detracts from the beauty of analytical and variational formulations. Nevertheless, recent advancements in applying variational principles in aerodynamics \citep{gonzalez2022variational} are promising, and encourages the current study.	

There are numerous efforts that developed variational principles of Euler's inviscid dynamics \citep{seliger1968variational,bateman1929notes,salmon1988hamiltonian,herivel1955derivation,bretherton1970note,penfield1966hamilton,hargreaves1908xxxvii,serrin1959mathematical,luke1967variational,loffredo1989eulerian,morrison1998hamiltonian,berdichevsky_book}. The reader is referred to the thorough review articles of \cite{salmon1988hamiltonian} and \cite{morrison1998hamiltonian}. Also, there are many efforts that aimed at extending these variational formulations to account for dissipative/viscous forces \citep{yasue1983variational,kerswell1999variational,gomes2005variational,eyink2010stochastic,fukagawa2012variational,galley2014principle,gay2017lagrangian}. In the current study, we mainly rely on the variational formulation of Euler's equation developed by \cite{seliger1968variational} using the principle of least action.

The principle of least action is typically stated as \citep{goldstein2002classical} \textit{``The motion of the system from time $t_1$ to time $t_2$ is such that the action integral $J$ is stationary''}
\begin{equation}\label{eq:action integral}
	J=\int_{t_1}^{t_2} \mathcal{L} (q_i, \dot{q}_i,t) \; dt,
\end{equation}
where $\mathcal{L}$ is the Lagrangian function, defined as $\mathcal{L}=T-V$ where $T$ and $V$ are the kinetic and potential energies, respectively, and $q_i$'s are the system generalized co-ordinates. A necessary condition for the functional $J$ to be stationary is that its first variation must vanish: $\delta J=0$ which, after applying calculus of variation techniques \citep{burns2013introduction}, results in the well-known Euler-Lagrange equation
\begin{equation}\label{eq:E-L}
	\frac{d}{dt}\left(\frac{\partial \mathcal{L}}{\partial \dot{q}_i}\right)-\frac{\partial \mathcal{L}}{\partial q_i}=0,\qquad i=1,2,....n.
\end{equation}
Noticeably, the representation in \cref{eq:E-L} describes the dynamics of  discrete particles only. However, if a continuum of particles is considered instead, the action integral $J$ is written in terms of a \textit{``Lagrangian Density''} $\mathcal{L}_d$ as
\begin{equation}\label{eq:first variation field}
	J=\int_{t_1}^{t_2} \int_{\Omega} \mathcal{L}_d\left(\bm{q},\bm{\dot{q}},\bm{q_{x}},\bm{x},t\right) \; d\Omega dt,
\end{equation}
where $\bm{x}$ are the spatial coordinate variables and $\Omega$ is the spatial domain. In this case, the generalized coordinates $\bm{q}$ are field variables (i.e., $\bm{q}=\bm{q}(\bm{x},t)$).


\subsection{Variational Formulation of Euler's Inviscid Dynamics: The Lagrangian Density is the Pressure}
The straightforward definition of the action integral for a fluid continuum
\begin{equation}\label{eq:PLA field unconstrained}
	J=\int_{t_1}^{t_2} \int_{\Omega} \left[\frac{1}{2} \rho \bm{u}^2 - \rho E(\rho,S)\right] \; d\Omega dt,
\end{equation}
where the first term in the integrand represents the KE, $E$ is the internal energy per unit mass, which represents the potential energy of the fluid continuum, and $S$ is the entropy whose changes are related to those of $E$ as
\begin{equation}\label{eq:energy thermodynamic}
	dE=\Theta dS-pd(1/\rho),
\end{equation}
where $p$ is the pressure, and $\Theta$ is the temperature. Starting with the action (\ref{eq:PLA field unconstrained}), through a long and rigorous proof that makes use of Clebsch representation and Lin's constraints, \cite{seliger1968variational} managed to show that the functional $J$ can be written in the Eulerian formulation as 
\begin{equation}\label{eq:Lagrnagina_density}
	J = - \int \int p\; d\Omega dt.
\end{equation}
That is the vanishing of first variation of $J$ yields the conservation equations of an inviscid fluid
\begin{subequations}\label{eq:cons laws}
	\begin{equation}\label{eq:Cont}
		\frac{\partial \rho}{\partial t}+ \nabla\cdot(\rho \bm{u})=0
	\end{equation}
	\begin{equation}\label{eq:mom}
		\rho \frac{D\bm{u}}{D t}=-\nabla p
	\end{equation}
	\begin{equation}\label{eq:energy}
		\frac{DS}{Dt}=0
	\end{equation}	
\end{subequations}
Hence, equation (\ref{eq:Lagrnagina_density}) implies that \textit{the Lagrangian density of a continuum of inviscid fluid is the pressure!}

Since the flow outside vortex patches is irrotational by definition, one can use the unsteady Bernoulli equation to write the pressure in terms of the velocity potential $\phi$. As such, the principle of least action will then imply that the first variation of the functional 
\begin{equation}\label{eq:PLA incomp irr}
	J= \int_{t_1}^{t_2} \int_{\Omega}  \rho \left\{\partial_t\phi+ \frac{1}{2} \left(\nabla\phi \right)^2\right\} \; d\Omega dt,
\end{equation}
must vanish. This statement is the cornerstone in our analysis below for vortex dynamics.

\section{Variational Dynamics of Rankine Vortex Patches as an Application}\label{sec:Variational Dynamics of Rankine Vortex Patches as an Application}
This section presents the core of this work where we utilize the proposed variational formulation to develop a novel model of vortex dynamics. In this application, $N$ free, circular, Rankine vortices are considered. In our formulation, each vortex must have a finite core (of radius $a_i$) to ensure a finite KE and allow the vortex inertial effects to appear. It is important to mention that constant-strengths and radii sizes are considered for simplicity and facilitating the comparison with KR model. However, the formulation is generic, allowing for unsteady vortices as well as deforming boundaries.

The domain $\Omega$ consists of an irrotational flow outside the vortex core and fluid under rigid body motion inside the core. Hence, knowing the vortices strengths a priori, the flow velocity $\bm{u}$ at any point $(x,y)$ in the domain can be written in terms of the locations $(x_i,y_i)$ of the vortices and their derivatives
\begin{equation}\label{eq:Vel_field_in_Terms_generalized_coordinates}
	\bm{u}(x,y)=\bm{u}(x_i,y_i,\dot{x}_i,\dot{y}_i;\Gamma_i).
\end{equation}
This representation is the foundation of the current analysis; the flow velocity in the entire domain depends only on a finite number of variables $(x_i,y_i)$. That is, \cref{eq:Vel_field_in_Terms_generalized_coordinates} represents a model reduction from infinite number of degrees of freedom down to only 2N. As such, the principle of least action, and analytical mechanics formulation in general, are especially well suited for this problem because it allows one to accept the given kinematical constraints (\ref{eq:Vel_field_in_Terms_generalized_coordinates}) and focus on the dynamics of the reduced system (2N degrees of freedom), in contrast to the Newtonian-mechanics formulation where such a reduction is not possible. Rather, the large system must be retained (i.e., a PDE for the infinite system at hand) and the kinematic constraint (\ref{eq:Vel_field_in_Terms_generalized_coordinates}) will be associated with an unknown constraint force in the equations of motion of the large system (in the PDE).

\begin{figure}
	\centering
	\includegraphics[keepaspectratio,width=3in]{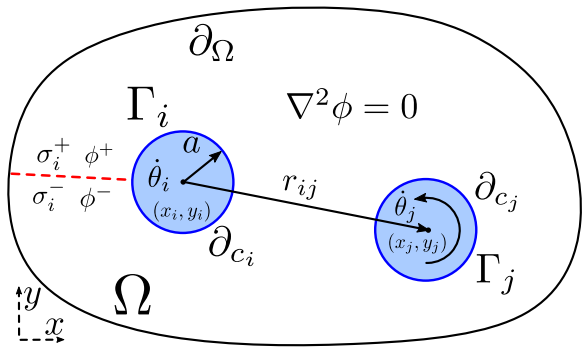}
	\caption{Schematic drawing for the problem; $\partial_j$ is the vortices boundaries (in blue), constructed barriers $\sigma_j$ (dashed red line) and flow filed domain boundary $\partial_{\Omega}$ (in black).}
	\label{fig:Final_representation_for_circular_vortices_presented_in_dept_sing_inkscape_JFM}
\end{figure}

The above description implies that the Lagrangian has two contributions
\begin{equation}\label{eq:Lag_flow+Lag_core}
	\mathcal{L}=\mathcal{L}_{\phi}+\mathcal{L}_C,
\end{equation}
where $\mathcal{L}_{\phi}$ is the Lagrangian of the irrotational flow outside the cores and $\mathcal{L}_C$ is the sum of the Lagrangians of the  fluid inside the cores, which is going through a rigid body motion. That is, part of the problem is a rigid body motion while the other is a continuum  field. In other words, the action integral for this problem formulation is given by
\begin{equation}\label{eq:action integral total}
	J=\int_{t_1}^{t_2} \left( \int_{\Omega} \mathcal{L}_{d_{\phi}} \; d\Omega + \mathcal{L}_C \right) (x_i(t),y_i(t),\dot{x}_i(t),\dot{y}_i(t);\Gamma_i) \; dt,
\end{equation}
where $\mathcal{L}_{d_\phi}$ is the Lagrangian density of the continuum field, which is $-p$ as shown by \cite{seliger1968variational} and presented in the previous section. Hence, the fluid Lagrangian density $\mathcal{L}_{d_\phi}$ will be integrated over space, then the problem could be treated as one with finite degrees of freedom $(x_i,y_i)$---akin to particle mechanics. Hence, there is only one independent variable (time) and the standard Euler-Lagrange equation (\ref{eq:E-L}) applies.

The Lagrangian of the core flow under rigid-body motion can be easily calculated as
\begin{equation}\label{eq:Rigid body Lag}
	\mathcal{L}_C=\sum\frac{m_i}{2}\bm{u}_i\cdot\bm{u}_i+\sum\frac{I_i}{2}\dot{\theta_i}^2,
\end{equation}
assuming the motion taking place in a horizontal plane (i.e., no active gravitational forces), where $m_i$ is the fluid mass inside the $i^{\rm{th}}$ vortex core and $I_i$ is its moment of inertia. As mentioned before, constant-strength vortices are considered, hence, the angular velocity $\dot{\theta}_i$ of each rigid core (which is proportional to $\Gamma_i$) is constant. This immediately forces the $\theta$ co-ordinate to be an ignorable/cyclic coordinate \citep{lanczos2020variational}. In other words, the corresponding momentum $\partial\mathcal{L}_C / \partial\dot{\theta}$ is constant. It now remains to compute the Lagrangian of the continuum fluid outside the cores in terms of the generalized coordinates $\bm{q}=(x_1,y_1,...,x_N,y_N)$.

\subsection{Irrotational Flow Lagrangian Computation}\label{sec:Flow lag}
As shown by \cite{seliger1968variational}, the Lagrangian density of an inviscid fluid is simply the pressure (actually $-p$). Hence, for the irrotational flow outside the cores, the Lagrangian is given by \cref{eq:PLA incomp irr}, where the second term is the flow KE and the first term is due to $\partial_t\phi$.

\subsubsection{Flow KE}\label{sec:flow KE}
Calculating the KE for vortex flows is not a trivial task. In fact, \cite{saffman} wrote \textit{``The use of variational methods for the calculation of equilibrium configurations is rendered difficult by the absence of simple and accurate formulae for the calculation of the kinetic energy''}. Overcoming this obstacle is one of the goals of this papers by deriving an accurate formula for a finite KE for vortex flows from basic definitions of mechanics.

Consider the flow KE of the irrotational flow field in \cref{fig:Final_representation_for_circular_vortices_presented_in_dept_sing_inkscape_JFM} 
\begin{equation}\label{eq:KE plain}
	T=\frac{\rho}{2}\int \bm{u}\cdot \bm{u} \; dA=\frac{\rho}{2}\int \left(\nabla\phi \right)^2 \; dA,
\end{equation}
where $A$ is the irrotational flow field area in the domain $\Omega$. After following the analysis in \cref{app:flow KE}, the KE could be written as: 
\begin{equation}\label{eq:KE final generic}
	T=\frac{\rho}{2}\biggr[\underset{i\neq j}{\sum \sum}\,\Gamma_i\psi_j|_{i}+\sum \Gamma_i\psi_i|_{\partial_{c_i}}+\Psi|_{R}\Gamma_R+\sum \Gamma_i\Psi|_{R}\biggr],
\end{equation}
where $\psi_j|_{i}$ is the stream function of the irrotational flow induced by the $j^{\rm{th}}$ vortex evaluated at the center of the $i^{\rm{th}}$ one, $\psi_i|_{\partial_{c_i}}$ is the $i^{\rm{th}}$ vortex stream function evaluated over its boundary $\partial_{c_i}$, $\Psi$ is the total stream function (due to all vortices; i.e., $\Psi=\sum_j \psi_j$), $\partial_{\Omega}$ is the outer boundary of the domain $\Omega$ and $\Gamma_R$ is the circulation over that boundary enclosing the whole domain which, for an irrotational flow, is simply given as the sum of all circulations inside (i.e., $\Gamma_R=\sum_i \Gamma_i$). The fourth term results from the multi-valued nature of the potential function $\phi$ when evaluated over the fictitious barrier $\sigma_i$ for the $i^{\rm{th}}$ vortex. Equation (\ref{eq:KE final generic}) is a generic result, but for the current study, we consider circular Rankine vortices of core radius $a$. Hence, the KE is written as 
\begin{equation}\label{eq:KE_final}
	T=-\frac{\rho}{4\pi}\underset{i\neq j}{\sum \sum}\,\Gamma_i\Gamma_j\ln r_{ij}-\frac{\rho}{4\pi}\left(\sum \Gamma_i^2\right)\ln a+\frac{\rho}{4\pi}\left(\sum \Gamma_i\right)^2\ln R+\Psi|_{R}\Gamma_R.
\end{equation}
These terms could be compared to the previously mentioned regularized KE presented in \cref{eq:regularized_ke}. It is clearly seen that the first term is exactly the KR function, while the second term in both equations are matching, with the core radius $a$ taking the place of the limit circle radius $\epsilon$, which avoids the unbounded KE because we have a finite core in contrast to $\epsilon\to 0$. So, this term will be retained, unlike the traditional analysis, which ignores it on the account that it is a constant term that should not affect the variational analysis. In fact, this term will be of paramount importance if $\Gamma_i$ is a degree of freedom, so its dynamics will be determined simultaneously from the same variational formulation. Also, the third term matches that in \cref{eq:regularized_ke}, which will blow up for unbounded flows except if the total circulation is zero. The fourth term is an additional term that is not captured in \cref{eq:regularized_ke}. However, similar to the third term, it is infinite for unbounded flows unless $\Gamma_R=0$. In fact, the main reason behind the boundedness of the last two terms in the case of zero total circulation is that, the velocity will be of order $\sim \mathcal{O}(1/r^2)$ in contrast to  $\sim \mathcal{O}(1/r)$ for a non-zero total circulation \citep{vortex_methods}. In summary, to obtain a finite KE, we must have: (1) vortex patches of finite sizes instead of infinitesimal ones (i.e., point vortices), and (2) zero total circulation. This last condition is interesting and thought provoking. Aside from the mathematical reason mentioned above ($\mathcal{O}(1/r^2)$ Vs $\mathcal{O}(1/r)$), we may think of the following physical reason. In reality, there should not be an unbounded irrotational flow with a non-zero total circulation. If we always start from a stationary state (i.e., $\Gamma_R=0$), Kelvin's conservation of circulation dictates that the total circulation must remain unchanged; if there is a circulation created somewhere in the flow (e.g., around a body), there must be a vortex of equal strength and opposite direction somewhere else in the flow field (e.g., a starting vortex over an airfoil \citep{glauertbook}). It is interesting that such a behavior is encoded into the KE.

It is important to mention that the evaluation of the previous KE holds irrespective of the vortex boundary shape, whether it is circular or elliptical or an arbitrary shape. However, if the boundary changes, it will result in a time-varying moment of inertia $I(t)$, precluding the cyclic nature of the angular motion $\theta$ even for constant-strength vortices. In this study, however, we ignore such changes in the vortex boundary shape.  This is a reasonable assumption as long as the distance between the vortices is considerably larger than the vortex core size \citep{lamb}. In other words, vortex patches could interact and deform when they are close to each other; this  behavior could be studied as a \textit{``Contour Dynamics''} problem \citep{zabusky1979contour,pullin1992contour_anual_review,saffman}. However, in this first-order study, we are concerned with vortex patches concentrated in tiny cores and hence ignore the effect of  shape deformation, which is one step beyond the point-vortex model. Luckily, it was shown by \cite{deem1978vortex,pierrehumbert1980family} and \cite{saffman1980equilibrium_shapes_pair_vortex_patches} that any vortex pairs distantly apart from each other, as long as $a/x_c<<1$ where $x_c$ is half the distance between the two vortices center, will behave as if they were point vortices and not deform the boundary of each other.

\subsubsection{$\partial_t \phi$ Contribution to the Lagrangian}\label{sec:partial phi partial t}
The flow total potential function is defined as
\begin{equation}\label{eq:flow potential}
	\phi=\sum_{i}\phi_i,
\end{equation}
where $\phi_i$ is the potential function of the flow associated with the $i^{\rm{th}}$ vortex, which depends on $(x,y,;x_i(t);y_i(t);\Gamma_i)$. That is, it does not have an explicit time-dependence. Rather, its time-derivative will be a convective-like term with respect to the Lagrangian co-ordinates $(x_i,y_i)$ whose velocity is $\bm{u}_i$
\begin{equation}\label{eq:partial_phi_as_gradient_flux}
	\partial_t\phi=\sum_{i}\bm{u}_i\cdot\nabla_i\phi_i.
\end{equation}
As such, the $\partial_t \phi$ contribution to the Lagrangian could be written as
\begin{equation}\label{eq:Lag_pphi_general}
	\mathcal{L}_{\phi_t}=\rho \int \partial_t \phi d\bm{x} = \rho\int\sum_{i}\bm{u}_i\cdot\nabla_i\phi_i\;dA,
\end{equation}
whose direct computation is cumbersome. Instead, we will apply \cref{eq:E-L} first. For example, consider the term $\frac{\partial \mathcal{L}_{\phi_t}}{\partial \dot{x}_i}$, which is needed for the $x_i$ equation of motion. By definition, we have
\begin{equation}\label{eq:pLphi_pxid}
	\frac{\partial \mathcal{L}_{\phi_t}}{\partial \dot{x}_i}=\rho\partial_{\dot{x}_i}\int\sum_{j}\bm{u}_j\cdot\nabla_j\phi_j\;dA,
\end{equation}
where the differentiation could be pushed forward within the integral because the integral bounds are independent of the vortex velocity $\dot{x}_i$. As such, all terms will vanish except 
\begin{equation}\label{eq:pLphi_pxid_simplified}
	\frac{\partial \mathcal{L}_{\phi_t}}{\partial \dot{x}_i}=\rho\int \partial_{x_i}\phi_i\;dA.
\end{equation}
Moreover, since there is no explicit time-dependence, assuming constant-strength vortices, the time derivative of the previous term is then given by
\begin{multline}\label{eq:d_dt_pLphi_pxid_simplified}
	\frac{d}{dt}\left(\frac{\partial \mathcal{L}_{\phi_t}}{\partial \dot{x}_i}\right)=\rho\bm{u}_i\cdot\nabla_i\left(\int \partial_{x_i}\phi_i\;dA\right)\\ = \rho\left\{\dot{x}_i\partial_{x_i}\int \partial_{x_i}\phi_i\;dA+\dot{y}_i\partial_{y_i}\int \partial_{x_i}\phi_i\;dA\right\}.
\end{multline}

Similarly, the other derivative $\partial \mathcal{L}_{\phi_t}/\partial x_i$ needed for the $x_i$ equation is written as
\begin{equation}\label{eq:pLphi_pxi}
	\frac{\partial \mathcal{L}_{\phi_t}}{\partial x_i}=\rho\partial_{x_i}\int\sum_{j}\bm{u}_j\cdot\nabla_j\phi_j\;dA=\rho\left\{\dot{x}_i\partial_{x_i}\int \partial_{x_i}\phi_i\;dA+\dot{y}_i\partial_{x_i}\int \partial_{y_i}\phi_i\;dA\right\}.
\end{equation}
At this moment, it should be obvious that subtracting \cref{eq:pLphi_pxi} from  \cref{eq:d_dt_pLphi_pxid_simplified} for the Euler-Lagrange equation of motion for $x_i$ will result in the cancellation of the first term. As such, we have
\begin{equation}\label{eq:xi_eq_pphi}
	\frac{d}{dt}\left(\frac{\partial \mathcal{L}_{\phi_t}}{\partial \dot{x}_i}\right) - \frac{\partial \mathcal{L}_{\phi_t}}{\partial x_i} = \rho\dot{y}_i\left[\partial_{y_i}\int \partial_{x_i}\phi_i\;dA-\partial_{x_i}\int \partial_{y_i}\phi_i\;dA\right],
\end{equation}
which, after the detailed computation presented in \cref{app:pphi}, yields 
\begin{equation}\label{eq:xi_yi_eq_pphi_final}
	\frac{d}{dt}\left(\frac{\partial \mathcal{L}_{\phi_t}}{\partial \dot{x}_i}\right) - \frac{\partial \mathcal{L}_{\phi_t}}{\partial x_i}=	\rho\dot{y}_i\biggr[\underset{i\neq j}{\sum_{j}} \left(\int_{\partial_{c_j}} \partial_{x_i}\phi_i\;dx-\int_{\partial_{c_j}} \partial_{y_i}\phi_i\;dy\right)-\Gamma_i\biggr],
\end{equation}
where the final integrals are performed over the  $j^{\rm{th}}$ vortex boundary ${\partial_{c_j}}$ and they are evaluated numerically. Also, for clarity and brevity, the numerical integrals are denoted as
\begin{equation}\label{eq:numerical_int_naming}
	I_{x_{ij}}=\int_{\partial_{c_j}} \partial_{x_i}\phi_i\;dx, \qquad I_{y_{ij}}=\int_{\partial_{c_j}} \partial_{y_i}\phi_i\;dy,
\end{equation}
resulting in the final form
\begin{equation}\label{eq:xi_eq_pphi_final_shortened}
	\frac{d}{dt}\left(\frac{\partial \mathcal{L}_{\phi_t}}{\partial \dot{x}_i}\right)- \frac{\partial \mathcal{L}_{\phi_t}}{\partial x_i}=	\rho\dot{y}_i\biggr[\underset{i\neq j}{\sum_{j}}\left(I_{x_{ij}}-I_{y_{ij}}\right)-\Gamma_i\biggr].
\end{equation}
Similarly, the $y_i$ equation is written as 
\begin{equation}\label{eq:yi_eq_pphi_final_shortened}
	\frac{d}{dt}\left(\frac{\partial \mathcal{L}_{\phi_t}}{\partial \dot{y}_i}\right) - \frac{\partial \mathcal{L}_{\phi_t}}{\partial y_i}=	\rho\dot{x}_i\biggr[\underset{i\neq j}{\sum_{j}}\left(I_{y_{ij}}-I_{x_{ij}}\right)+\Gamma_i\biggr].
\end{equation}

\subsection{Equations of Motion}
We are now ready to write the final form of the equations of motion from the Euler-Lagrange equation (\ref{eq:E-L}), where $\bm{q}=(x_1,y_1,...,x_N,y_N)$, the Lagrangian $\mathcal{L}=\mathcal{L}_{\phi} + \mathcal{L}_C$, the contributions of different terms are given in \cref{eq:KE_final,eq:Rigid body Lag,eq:xi_eq_pphi_final_shortened,eq:yi_eq_pphi_final_shortened}.

Substituting \cref{eq:KE_final,eq:Rigid body Lag,eq:xi_eq_pphi_final_shortened,eq:yi_eq_pphi_final_shortened} into the Euler-Lagrange equation (\ref{eq:E-L}), while assuming the motion taking place in a horizontal plan, we obtain
\begin{align}\label{eq:xi_yi_eq_T_plus_core_total}
	m_i\ddot{x}_i+\frac{\rho\Gamma_i}{2\pi}\underset{i\neq j}{\sum_{j}}\,\Gamma_j\frac{(x_i-x_j)}{r_{ij}^2}+\rho\dot{y}_i\biggr[\underset{i\neq j}{\sum_{j}}\left(I_{x_{ij}}-I_{y_{ij}}\right)-\Gamma_i\biggr]=0\\
	m_i\ddot{y}_i+\frac{\rho\Gamma_i}{2\pi}\underset{i\neq j}{\sum_{j}}\,\Gamma_j\frac{(y_i-y_j)}{r_{ij}^2}+\rho\dot{x}_i\biggr[\underset{i\neq j}{\sum_{j}}\left(I_{y_{ij}}-I_{x_{ij}}\right)+\Gamma_i\biggr]=0
\end{align}
which considering $m_i=\rho\pi a^2$, results in
\begin{equation}\label{eq:xi_eq_T_plus_core_total_simplified}
	\ddot{x}_i=-\frac{\Gamma_i}{2\pi^2a^2}\underset{i\neq j}{\sum_{j}}\,\Gamma_j\frac{(x_i-x_j)}{r_{ij}^2}-\frac{\dot{y}_i}{\pi a^2}\biggr[\underset{i\neq j}{\sum_{j}}\left(I_{x_{ij}}-I_{y_{ij}}\right)-\Gamma_i\biggr]
\end{equation}
\begin{equation}\label{eq:yi_eq_T_plus_core_total_simplified}
	\ddot{y}_i=-\frac{\Gamma_i}{2\pi^2a^2}\underset{i\neq j}{\sum_{j}}\,\Gamma_j\frac{(y_i-y_j)}{r_{ij}^2}-\frac{\dot{x}_i}{\pi a^2}\biggr[\underset{i\neq j}{\sum_{j}}\left(I_{y_{ij}}-I_{x_{ij}}\right)+\Gamma_i\biggr]
\end{equation}
These equations represent the sought dynamical equations of motion that describe true dynamics of free vortices from first principles: the Principle of Least Action. They are second-order in nature, capturing inertial effects of the core. Interestingly, in the limit of a vanishing core size ($a\rightarrow0$), $m_i$, $I_{x_{ij}}$ and $I_{y_{ij}}$ go to zero, and the resulting dynamics \cref{eq:xi_eq_T_plus_core_total_simplified,eq:yi_eq_T_plus_core_total_simplified} reduce to the first-order equations given by the KR Hamiltonian. However, for a finite-size cores, the new dynamics given by \cref{eq:xi_eq_T_plus_core_total_simplified,eq:yi_eq_T_plus_core_total_simplified} are richer than the first-order kinematic equations of the KR Hamiltonian, which merely recover the Biot-Savart law. In particular, they allow enforcing an initial condition on velocity, similar to any typical problem in dynamics. Moreover, the analysis allows for extension to time-varying vortices where $\dot\Gamma$-terms will appear. Also, arbitrary conservative forces (e.g., gravitational, electric) can be easily incorporated in the framework, in contrast to the standard analysis based on the KR-function. These extensions will be considered in future work. However, to show the value of the proposed new formulation, some case studies will be presented below. Moreover, if dense-mass, point vortices are considered (i.e., the numerical integrals $I_{x_{ij}}$, $I_{y_{ij}}$ vanish, but the core mass $m_i$ is retained), the resulting equations of motion along with the Hamiltonian (see \cref{app:Hamiltonian_Ragazzo}) reduce to those deduced by \cite{ragazzodensevortex_electro_simmilarity} relying on analogy with electromagnetism. The recovery of the proposed formulation to the special cases of the KR formulation and Ragazzo's provides some credibility to the presented approach.

\begin{figure}
	\centering
	\begin{subfigure}{.45\textwidth}
		\centering
		\includegraphics[keepaspectratio,width=\textwidth]{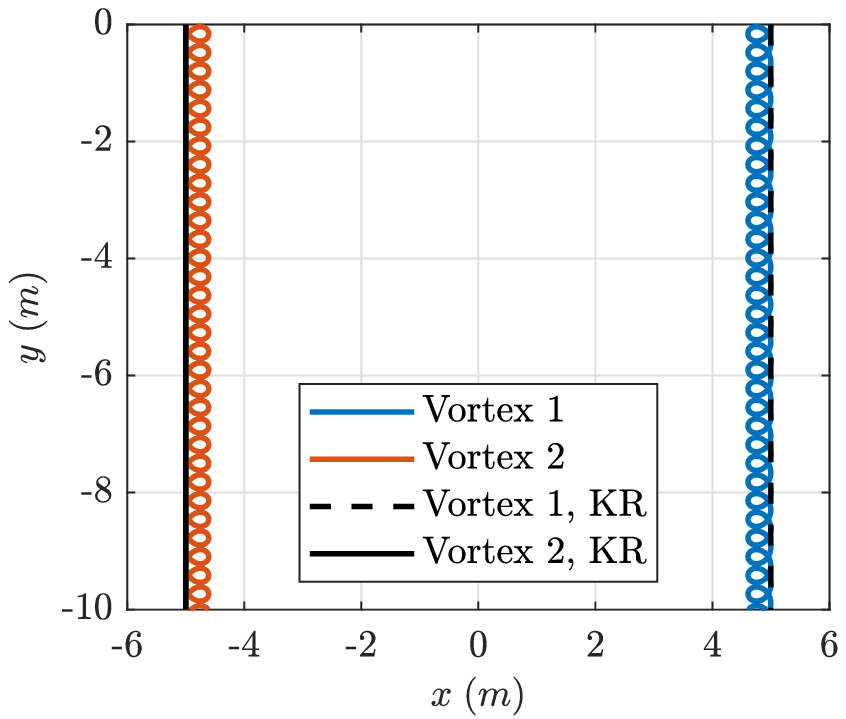}
		\caption{counter-rotating pair}
		\label{fig:a=1_G1=-G2=10pi_vo=5vi_t=20_VS_KR}
	\end{subfigure}
	\begin{subfigure}{.45\textwidth}
		\centering
		\includegraphics[keepaspectratio,width=\textwidth]{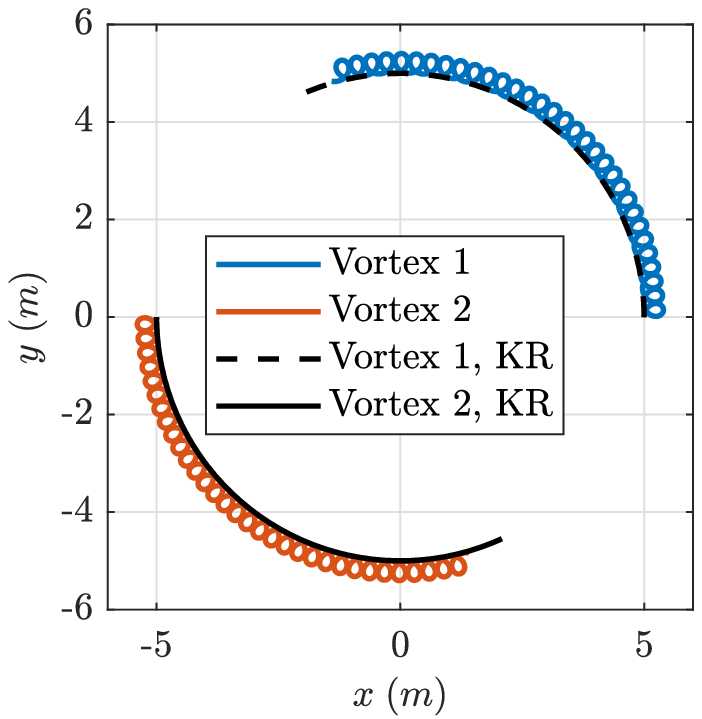}
		\caption{co-rotating pair}
		\label{fig:a=1_G1=G2=10pi_vo=5vi_t=20_VS_KR}
	\end{subfigure}
	\caption{Simulation of proposed variational vortex dynamics model against the KR model for a pair of vortices. The initial velocities are set to five times the Biot-Savart induced velocity in the same direction.}
	\label{fig:PLA_VS_KR}
\end{figure}

\section{Case Studies}\label{sec:new_results}
Different case studies are presented to highlight the similarity and difference between the motions resulting from the proposed model and the standard KR model (i.e., Biot-Savart law). As pointed above, the proposed variational dynamics result in the KR solution for small vortex radii. However, one of the major differences between the two formulations is that the kinematic KR equations are first-order allowing only initial conditions of vortex position to be assigned whereas the proposed equations are second-order, admitting arbitrary initial velocities in addition. Therefore, the similarity between the resulting two motions does not necessarily happen  when the initial velocities do not match the induced velocities by the  Biot-Savart law.

Equal strength vortex couples are considered in \cref{fig:PLA_VS_KR} with radius $a=1$ and $\Gamma=10\pi$. The trajectories of the vortex couples are presented in comparison to the KR motion. The trajectory of a counter-rotating vortex couple is shown in \cref{fig:a=1_G1=-G2=10pi_vo=5vi_t=20_VS_KR}  while \cref{fig:a=1_G1=G2=10pi_vo=5vi_t=20_VS_KR} shows the trajectory of a co-rotating pair. For this case of small vortex cores (in comparison to the relative distance), when the model is initialized with the Biot-Savart induced velocity, almost-exact matching with the KR motion is obtained (black trajectory in \cref{fig:PLA_VS_KR}). However, when the proposed model is initialized with a different value in the same direction, the resulting dynamics (red and blue trajectories in \cref{fig:PLA_VS_KR}) are richer. The resulting motion is composed of fast and slow time scales. The slow dynamics (average solution) possesses a similar behavior to the KR one (in many but not all cases, as shown below), though the vortex pairs act like magnets; opposite sign attract (\cref{fig:a=1_G1=-G2=10pi_vo=5vi_t=20_VS_KR}), and same sign repel (\cref{fig:a=1_G1=G2=10pi_vo=5vi_t=20_VS_KR}). Around this averaged response, there are oscillations in the fast time scale.

\begin{figure}
	\centering
	\includegraphics[keepaspectratio,width=0.4\textwidth]{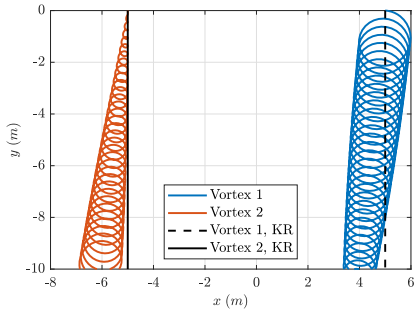}
	\caption{Simulation of the variational vortex dynamics model presenting the deviation of a counter-rotating vortex couple from the KR trajectory. Initial velocities for the vortices are $v_{1_0}=(0i,-0.5j)$ and  $v_{2_0}=(10i,-0.5j)$, respectively.}
	\label{fig:a=1_G1=-G2=10pi_vol=(0i_vij)_vor=(20vii_vij)=t=20_VS_KR}
\end{figure}

The behavior of the slow dynamics is not always similar to that of the KR model. In \cref{fig:a=1_G1=-G2=10pi_vol=(0i_vij)_vor=(20vii_vij)=t=20_VS_KR}, a pair of counter-rotating vortices are simulated, and the system is initialized with the Biot-Savart induced velocity (i.e., down), however, the vortex on the right is given an additional initial velocity to the right. These initial conditions resulted in quite  different averaged behaviors of the two vortices, as shown in \cref{fig:a=1_G1=-G2=10pi_vol=(0i_vij)_vor=(20vii_vij)=t=20_VS_KR}. The effect of the initial velocity on the averaged motion of interacting vortices, as demonstrated above, points to the interesting application of vortex motion control. Imagine a large vortex (e.g., a hurricane) that we like to deviate its motion from an anticipated trajectory. One can then pose the interesting question: Can we set a group of vortices in motion with specified initial velocities in the neighborhood of the large vortex so that its net motion deviates from the anticipated trajectory in the absence of these vortices (e.g., a hurricane misses hitting the land)? Using a similar approach to the one presented above, one can answer such a question.

To further demonstrate the capability of the developed vortex model to capture dynamical features that cannot be directly captured by the usual KR formulation, we consider a charged particle inside the core of each vortex and that the motion takes place in the presence of a constant-strength electric field. In this case, each vortex will experience a \textit{Lorentz} force $\bm{F}=q\bm{E}$, where $q$ is the particle's charge and $\bm{E}$ is the electric field vector. The effect of this electric field on the motion of the vortex system cannot be directly obtained from the standard kinematic formulation using the KR function. A true dynamical formulation is invoked instead. Therefore, it is straightforward to account for this effect using the developed dynamical formulation where forces can be considered. Simply, the right hand sides of \cref{eq:xi_eq_T_plus_core_total_simplified,eq:yi_eq_T_plus_core_total_simplified} will be modified by adding $q_i E_x/m_i$ and $q_i E_y/m_i$, respectively, where $E_x$ and $E_y$ are the components of the electric field in the $x$ and $y$ directions, respectively, and $q_i$ is the charge of the $i^{\rm{th}}$ vortex. Alternatively, one could add $-\sum_i q_i\Phi(x_i,y_i)$ to the Lagrangian function, where $\Phi$ is the scalar electrostatic potential ($\bm{E}=-\bm\nabla \Phi$). For simplicity, we considered a large electric field and small charges so that the Coulomb force is negligible with respect to the Lorentz force. In the following simulations, we considered $|q\bm{E}|/m=g$, where $g$ is the gravitational acceleration, and all motions are initialized with the KR velocities.

\begin{figure}
	\centering
	\begin{subfigure}{.32\textwidth}
		\centering
		\includegraphics[keepaspectratio,width=\textwidth]{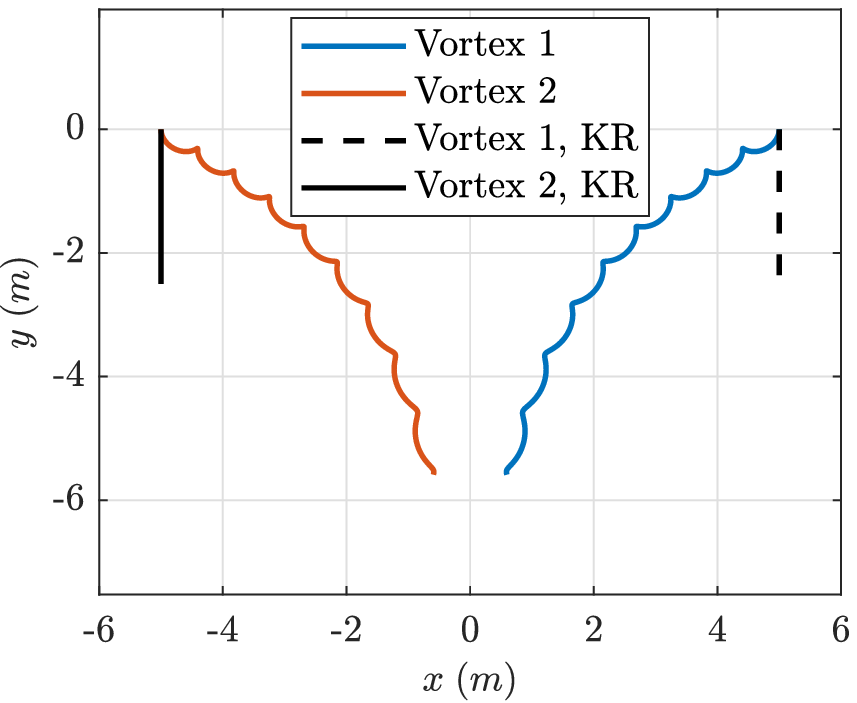}
		\caption{}
		\label{fig:a=1_G1=-G2=10pi_l=5_vo=vi_-g1_-g2}
	\end{subfigure}
	\begin{subfigure}{.32\textwidth}
		\centering
		\includegraphics[keepaspectratio,width=\textwidth]{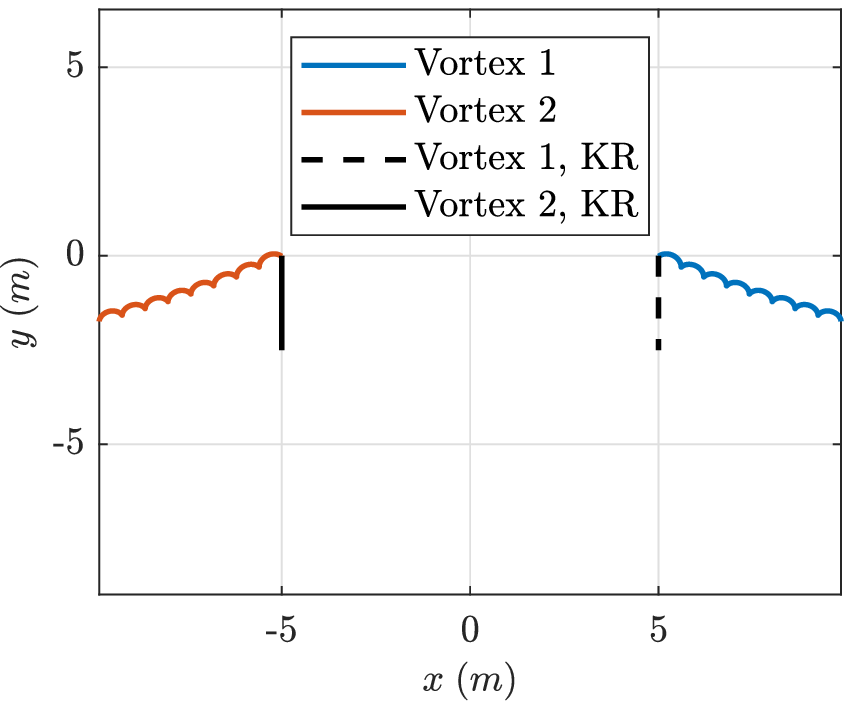}
		\caption{}
		\label{fig:a=1_G1=-G2=10pi_l=5_vo=vi_+g1_+g2}
	\end{subfigure}
	\begin{subfigure}{.32\textwidth}
		\centering
		\includegraphics[keepaspectratio,width=\textwidth]{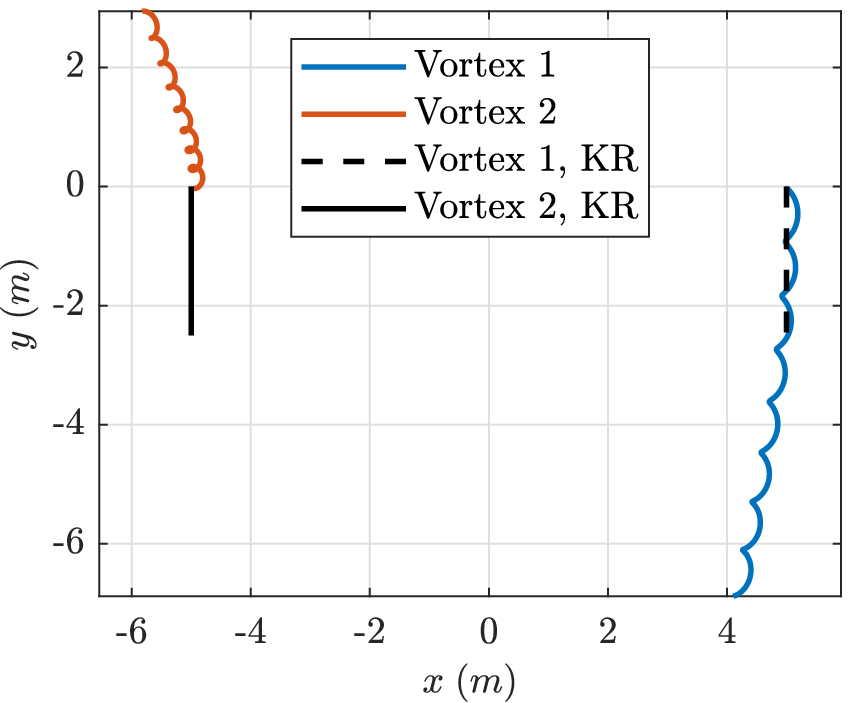}
		\caption{}
		\label{fig:a=1_G1=-G2=10pi_l=5_vo=vi_+xg1_+xg2}
	\end{subfigure}
	\caption{Different simulations for counter-rotating pair of vortices with same charge placed in a constant-strength electric field. Simulations are initialized by the Biot-Savart induced velocity and electric field direction for each case is listed as: (a) $\bm{E}\downarrow$, (b) $\bm{E}\uparrow$ and (c) $\bm{E}\rightarrow$.}
	\label{fig:Eelectric_gravity_in_same_direction}
\end{figure}

\begin{figure}
	\centering
	\begin{subfigure}{.32\textwidth}
		\centering
		\includegraphics[keepaspectratio,width=\textwidth]{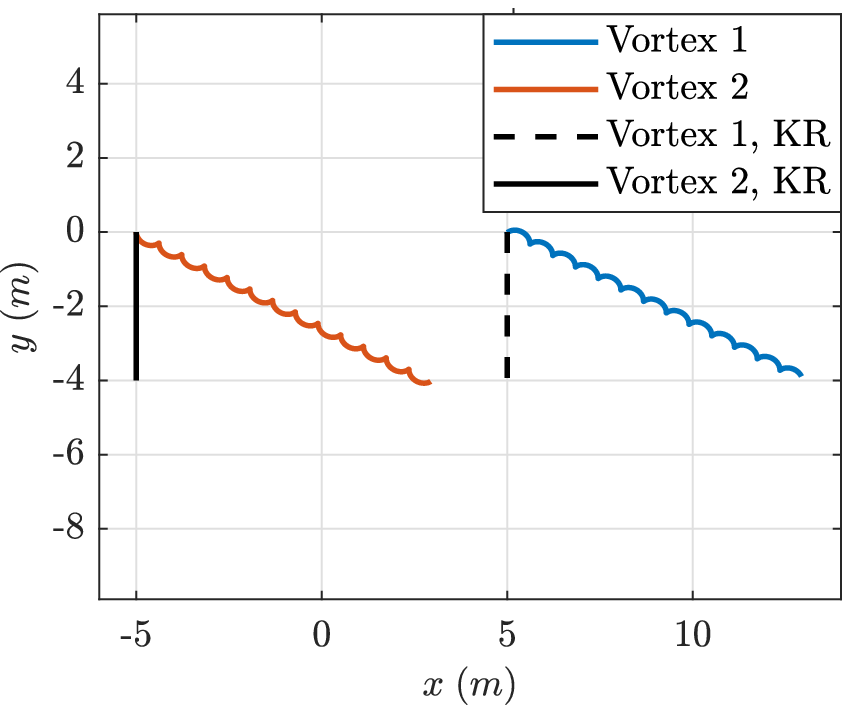}
		\caption{}
		\label{fig:a=1_G1=-G2=10pi_l=5_vo=vi_+g1_-g2}
	\end{subfigure}
	\begin{subfigure}{.32\textwidth}
		\centering
		\includegraphics[keepaspectratio,width=\textwidth]{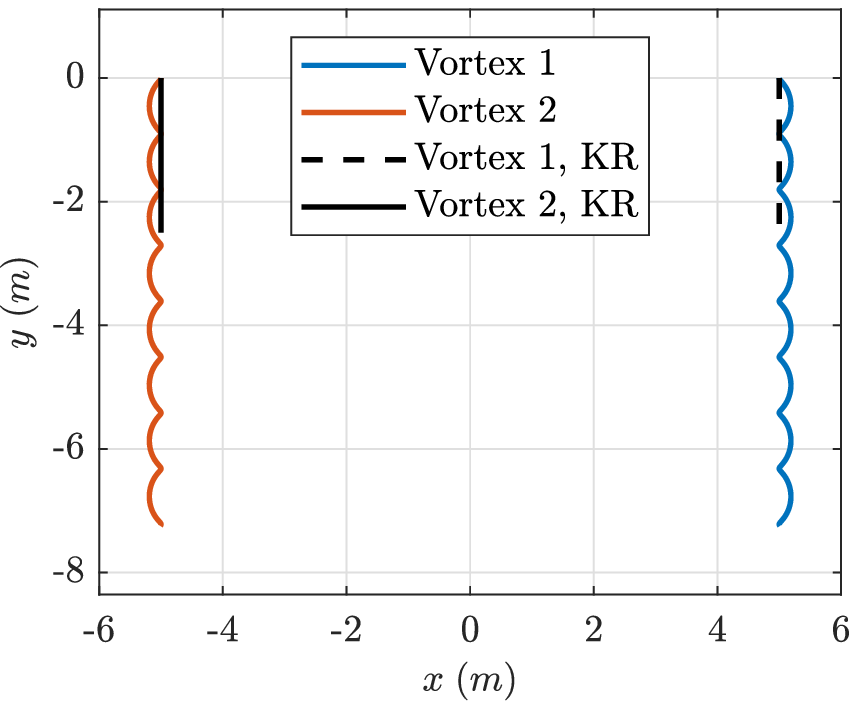}
		\caption{}
		\label{fig:a=1_G1=-G2=10pi_l=5_vo=vi_+xg1_-xg2}
	\end{subfigure}
	\begin{subfigure}{.32\textwidth}
		\centering
		\includegraphics[keepaspectratio,width=\textwidth]{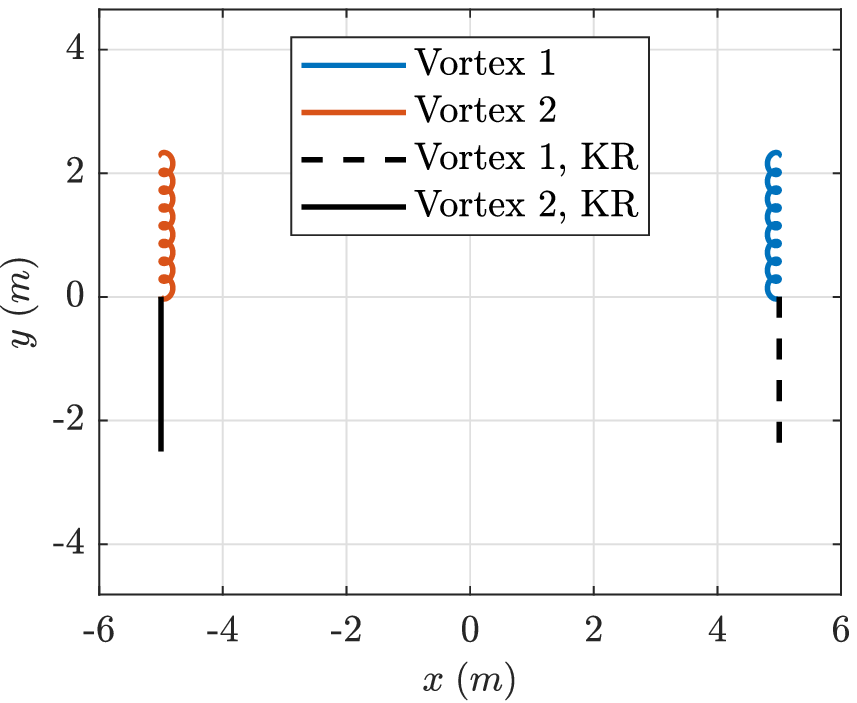}
		\caption{}
		\label{fig:a=1_G1=-G2=10pi_l=5_vo=vi_-xg1_+xg2}
	\end{subfigure}
	\caption{Different simulations for counter-rotating pair of vortices with opposite charge placed in a constant-strength electric field. Simulations are initialized by the Biot-Savart induced velocity; electric field direction and charge sign for each case are listed as: (a) $\bm{E}\uparrow$, $+q_1$ and $-q_2$, (b) $\bm{E}\rightarrow$, $+q_1$ and $-q_2$ and (c) $\bm{E}\rightarrow$, $-q_1$ and $+q_2$.}
	\label{fig:Eelectric_gravity_in_opposite_direction}
\end{figure}

\Cref{fig:Eelectric_gravity_in_same_direction} shows a pair of counter-rotating vortices of the same strength and the same charge placed in an electric field of constant strength. \Cref{fig:a=1_G1=-G2=10pi_l=5_vo=vi_-g1_-g2} shows the resulting motion when the electric field is pointing downward (negative $y$). This implies that the vortices will accelerate downward beyond the KR value dictated by the Biot-Savart law, which will in turn cause an acceleration in the $x$-direction because of the term $\dot{y}_i \Gamma_i$ in \cref{eq:xi_eq_T_plus_core_total_simplified}. As a result, the two counter-rotating vortices attract towards one other. Note that the simulation should not be deemed valid when the vortices get very close to each other. On the other hand, reversing the direction of the electric field (i.e., opposite to the KR velocity), the KR $y$-motion is decelerated, leading the counter-rotating vortices to repel, as shown in \cref{fig:a=1_G1=-G2=10pi_l=5_vo=vi_+g1_+g2}. Interestingly, applying an electric field in the horizontal direction ($+x$), we obtain the response in \cref{fig:a=1_G1=-G2=10pi_l=5_vo=vi_+xg1_+xg2}, which is non-intuitive. Although the applied electric force is in the $x$-direction, the net effect is much more significant in the $y$-direction. Moreover, even the effect in the $x$-motion is counter-intuitive; both vortices drift in the negative $x$-axis (opposite to the direction of the applied electric force). At the beginning, there is an acceleration for both vortices in the $x$-direction. However, the slightest velocity in $x$ activates the term $-\dot{x}_i \Gamma_i$ in the $\ddot{y}_i$ \cref{eq:yi_eq_T_plus_core_total_simplified}, which causes a downward acceleration for the right vortex and an upward acceleration for the left vortex (of negative strength). These vertical accelerations, in turn, affect the $x$-motion through the term $\dot{y}_i \Gamma_i$ in \cref{eq:xi_eq_T_plus_core_total_simplified}, causing both vortices to drift to the left (opposite to the applied force). This interesting interaction between the motion induced by the electric force and the motion induced by the \textit{vortex force} ($\dot{y}_i \Gamma_i$,-$\dot{x}_i \Gamma_i$) is naturally captured in the developed dynamic model. Although the KR motion is not really relevant here, it is presented for comparison in \cref{fig:Eelectric_gravity_in_same_direction,fig:Eelectric_gravity_in_opposite_direction}.

\Cref{fig:Eelectric_gravity_in_opposite_direction} shows the motion of a pair of counter-rotating vortices of equal strength, but of equal and opposite charges placed in a constant-strength electric field. \Cref{fig:a=1_G1=-G2=10pi_l=5_vo=vi_+g1_-g2} shows the response to an upward electric field causing an upward force on the right vortex (with positive charge) and a downward force on the left vortex (with negative charge). Yet, both vortices move downward and the significant effect is a drift in the $x$-direction due to the interaction mentioned above: the downward acceleration of the left vortex causes an acceleration in the $x$-direction through the term $\dot{y}_i \Gamma_i$ in \cref{eq:xi_eq_T_plus_core_total_simplified}, which in turn causes an upward acceleration through the term $-\dot{x}_i \Gamma_i$ in \cref{eq:yi_eq_T_plus_core_total_simplified}. Similar behavior occurs with the right vortex; initially, it experiences an  upward acceleration from the electric force, which causes an $x$-acceleration through the term $\dot{y}_i \Gamma_i$ in \cref{eq:xi_eq_T_plus_core_total_simplified}. This $x$-motion, in turn, causes a negative $y$-acceleration through the term $-\dot{x}_i \Gamma_i$ in \cref{eq:yi_eq_T_plus_core_total_simplified}. As a result, the right vortex, though forced by an upward force, moves to the right and downward, which is quite a non-intuitive behavior. 

\Cref{fig:a=1_G1=-G2=10pi_l=5_vo=vi_+xg1_-xg2} shows the response to an electric field in the $x$-direction, causing a right force on the right vortex and a left force on the left vortex (with the opposite charge); i.e., pulling the vortices away from each other. However, no considerable net effect is observed in contrast to the other scenarios. The initial right motion of the right vortex due to the electric force causes a downward acceleration, which in turn causes a negative $x$-acceleration that opposes the electrodynamic acceleration. The situation leads to an equilibrium with a periodic solution close to the KR motion. The behavior in \cref{fig:a=1_G1=-G2=10pi_l=5_vo=vi_-xg1_+xg2} is quite interesting, which presents the motion due to an electric field to the left, causing a left force on the right vortex and a right force on the left vortex; i.e., pushing the vortices towards one another. However, the response is counter-intuitive as usual due to the interesting interaction between the electrodynamic force and the hydrodynamic vortex force. Instead of getting closer to each other (to comply with the applied force), the two vortices move upward opposite to their initial KR velocity. As usual, the vortices initially follow the applied electric force (i.e., get closer to one another). This $x$ motion causes an upward acceleration to both vortices, which causes an $x$-acceleration opposite to the electrodynamic one for both vortices. 

The interaction between the electrodynamic force and the vortex force ($\dot{y}_i \Gamma_i$ in the $x$-direction and $-\dot{x}_i \Gamma_i$ in the $y$-direction) leads to very interesting behaviors: when pulling vortices downward (i.e., in the same direction as their KR initial velocity), they attract (\cref{fig:a=1_G1=-G2=10pi_l=5_vo=vi_-g1_-g2}); when pulling them upward (i.e., opposite to their KR initial velocity), they repel (\cref{fig:a=1_G1=-G2=10pi_l=5_vo=vi_+g1_+g2}); when they pushed together to the right, they both end up moving to the left with one upward and one downward (\cref{fig:a=1_G1=-G2=10pi_l=5_vo=vi_+xg1_+xg2}); when pulling them away from each other, the almost did not respond (\cref{fig:a=1_G1=-G2=10pi_l=5_vo=vi_+xg1_-xg2}); and when pushing them towards one another, they move upward (\cref{fig:a=1_G1=-G2=10pi_l=5_vo=vi_-xg1_+xg2}). It is quite non-intuitive, yet explainable from the physics of the dynamic model in \cref{eq:xi_eq_T_plus_core_total_simplified,eq:yi_eq_T_plus_core_total_simplified}. This nonlinear behavior presents the dynamic model (\ref{eq:xi_eq_T_plus_core_total_simplified}, \ref{eq:yi_eq_T_plus_core_total_simplified}) as a rich mechanical system for geometric mechanics and control analysis using Lie brackets \citep{bullo2019geometric,hassan2019differential,taha2021geometric} where the concepts of \textit{anholonomy} \citep{Baillieul1996}, geometric phases \citep{marsden1991symmetry,marsden1997geometric}, and nonlinear controllability \citep{hassan2017geometric,taha2021nonlinear} can be demonstrated.

\section{Conclusion}
We introduced the use of Hamilton's principle of least action to develop a novel variational formulation for vortex dynamics. The developed model is fundamentally different from previous models in the literature, which are typically devised to recover a pre-known set of ODEs that are purely kinematical (first-order in position). In fact, they merely provide a Hamiltonian formulation of the kinematic equations of the Biot-Savart law. In contrast, the new model is based on formal variational principles for a continuum of inviscid fluid. As such, the model is dynamic in nature, constituting of second-order ODEs in position, which allows specifying initial velocities for the vortex patches and external forces (e.g., electromagnetic). The model provided rich and intriguing dynamics for counter and co-rotating vortex pairs. In the limit to a vanishing core size, the Biot-Savart kinematical behavior is recovered. For a finite core size, there is a multi-time-scale behavior: a slow dynamics that resembles the Biot-Savart motion, and a fast dynamics that results in fast oscillations around the Biot-Savart motion. However, for some given initial conditions, the averaged motion is considerably different from the Biot-Savart motion. The situation becomes interesting when an electric field is applied to a charged particle inside the vortex core. In this case, the interaction between the electrodynamic force and the hydrodynamic vortex force leads to non-intuitive behavior. For example, when a pair of counter-rotating vortices were pulled away from each other by the electric force, they moved normal to this force and opposite to their initial Biot-Savart motion; and when  they were pulled in the same direction of their initial velocity, they attract to one another (normal to the applied force). This interesting nonlinear dynamics present the developed dynamical model as a rich example in geometric mechanics and control where concepts of anholonomy, geometric phases, and nonlinear controllability can be demonstrated. 

\section*{Acknowledgments}
This material is based upon work supported by National Science Foundation grant number CBET-2005541. The first author would like to thank Mahmoud Abdelgalil, Ph.D., at University of California Irvine for fruitful discussions.


\bibliographystyle{jfm}
\bibliography{Vortex_PLA_literature}

\appendix
\section{Flow KE Integrals}\label{app:flow KE}

The KE calculation procedure is shown here, it relies on potential flow relations, topology and integral theorems. Reader knowledge are assumed to be present, for familiarization with these topics please refer to \cite{lamb,milnehydro,milneaero,batchelor,saffman,eldredge,daren_crowdy}. The 2D KE is given as 
\begin{equation}\label{eq:KE app}
	T=\frac{\rho}{2}\int \left(u^2+v^2\right) \;dA,
\end{equation}
knowing that 2D irrotational flow satisfies the following
\begin{equation}\label{eq:2D irr}
	u=\frac{\partial\psi}{\partial y}, \quad v=\frac{-\partial\psi}{\partial x}, \quad \omega_z=\frac{\partial v}{\partial x}-\frac{\partial u}{\partial y}=0,
\end{equation}
and substituting it in \cref{eq:KE app}, will result in 
\begin{equation}\label{eq:KE app 2}
	T=\frac{\rho}{2}\int \left(u\frac{\partial\psi}{\partial y}-v\frac{\partial\psi}{\partial x}-\psi\omega_z\right)\;dA=\frac{\rho}{2}\int \frac{\partial(u\psi)}{\partial y}-\frac{\partial( v\psi)}{\partial x} \;dA=\frac{\rho}{2}\int \psi\omega_z\;dA.
\end{equation}
This final representation of the KE  applies for irrotational and rotational flows, however, we will restrict the current computation for irrotational only. The area integrals will be converted to line ones using Stokes' theorem. This conversion is performed after transforming the multiply-connected domain to a simply-connected domain through the construction of fictitious barriers found in \cref{fig:Final_representation_for_circular_vortices_presented_in_dept_sing_inkscape_JFM}, which yields

\begin{equation}\label{eq:KE app 3}
	T=\frac{\rho}{2}\oint \psi \bm{u}\cdot \bm{dl} =\frac{\rho}{2}\int \psi d\phi.
\end{equation}
The above expression will be evaluated carefully over the different domain boundaries and constructed barriers while taking in consideration the cyclic constant $\Gamma$ of the multi-valued function $\phi$. The  result is represented as:
\begin{multline}\label{eq:KE app 4}
	T=\frac{\rho}{2}\left[\int_{\partial_{c_i}} \psi d\phi+\int_{\partial_{\Omega}}\psi d\phi+\int_{\sigma_i}\psi d\phi\right]\\=\frac{\rho}{2}\biggr[\underset{i\neq j}{\sum \sum}\,\Gamma_i\psi_j|_{i}+\sum \Gamma_i\psi_i|_{\partial_{c_i}}+\Psi|_{R}\Gamma_R+\sum \Gamma_i\Psi|_{R}\biggr].
\end{multline}
By this, the KE representation is achieved and further analysis over different terms with extra simplifications are discussed in \cref{sec:flow KE}.

\section{$\partial_t\phi$ Lagrangian Calculation}\label{app:pphi}

Starting from \cref{eq:xi_eq_pphi} the integrals will be evaluated to reach the final form in \cref{eq:xi_yi_eq_pphi_final}. Each term $I_y$ and $I_x$, will be computed separately as shown below.
\begin{equation}\label{eq:pphi_1}
	\frac{d}{dt}\left(\frac{\partial \mathcal{L}_{\phi_t}}{\partial \dot{x}_i}\right) - \frac{\partial \mathcal{L}_{\phi_t}}{\partial x_i} =	\rho\dot{y}_i\biggr[\underbrace{\partial_{y_i}\int \partial_{x_i}\phi_i \; dA}_{I_y}-\underbrace{\partial_{x_i}\int \partial_{y_i}\phi_i \; dA}_{I_x}\biggr].
\end{equation}
Relying on the reciprocal property of the potential function $\phi$ between parameterization $(x,y)$ co-ordinates and generalized one $(x_i,y_i)$,
\begin{equation}\label{eq:reciprocal}
	\partial_{y_i}\phi_i=-\partial_{y}\phi_i, \qquad \partial_{x_i}\phi_i=-\partial_{x}\phi_i,
\end{equation}
the $I_y$ integral could be simplified and transformed to a boundary integral as:
\begin{equation}\label{eq:I_y}
	I_y=\partial_{y_i}\int \partial_{x_i}\phi_i \; dA=-\partial_{y_i}\int \partial_{x}\phi_i \; dxdy=-\partial_{y_i}\int_\partial \phi_i \; dy.
\end{equation}
Following \cref{fig:Final_representation_for_circular_vortices_presented_in_dept_sing_inkscape_JFM} the integral will be performed for each boundary and barriers. Noticing that the barriers are taken to be parallel to the $x$ axis, then $I_y$ will be
\begin{equation}\label{eq:Iy_expanded}
	I_y=-\partial_{y_i}\biggl\{\underset{i\neq j}{\sum_{j}}\int_{\partial_{c_j}} \phi_i\;dy+\int_{\partial_{\Omega}} \phi_i\;dy+\int_{\sigma_i^+} \phi_i\;dy+\int_{\sigma_i^-} \phi_i\;dy+\int_{\partial_{c_i}} \phi_i\;dy\biggl\},
\end{equation}
treating $\phi_i$ as a single-valued function over the boundaries after constructing the barriers, and making sure that $R>>>>{x_i,y_i}$, then the last and second integrals will vanish. Moreover, the third and fourth integral will vanish, the former because of the barrier location and the latter because of setting $\phi_i^-=0$. By this, the first integral is the only one left and the differentiation could act over the integrand because the integral boundary $\partial_{c_j}$ is independent of $y_i$ (i.e., no need for Leibniz rule). The final form of $I_y$ integral will be
\begin{equation}\label{eq:Iy_final}
	I_y=-\underset{i\neq j}{\sum_{j}}\int_{\partial_{c_j}} \partial_{y_i} \phi_i\;dy,
\end{equation}
where it is evaluated numerically over the $j^{\text{th}}$ vortex boundary.

The $I_x$ integral will be computed following the same analysis, and it could be written as 
\begin{equation}\label{eq:Ix_expanded}
	I_x=-\partial_{x_i}\biggl\{\underset{i\neq j}{\sum_{j}}\int_{\partial_{c_j}} \phi_i\;dx+\int_{\partial_{\Omega}} \phi_i\;dx+\int_{\sigma_i^+} \phi_i\;dx+\int_{\sigma_i^-} \phi_i\;dx+\int_{\partial_{c_i}} \phi_i\;dx\biggl\}.
\end{equation}
Again as before, the second, fourth and last term vanishes with same analogy. However, the third term will not vanish as $\phi_i^+=2\pi$ and it will result into 
\begin{equation}\label{eq:Ix_simplified}
	I_x=-\partial_{x_i}\biggl\{\underset{i\neq j}{\sum_{j}}\int_{\partial_{c_j}}  \phi_i\;dx+\Gamma_i(R-x_i-a)   \biggl\}=-\underset{i\neq j}{\sum_{j}}\int_{\partial_{c_j}} \partial_{x_i} \phi_i\;dx+\Gamma_i.
\end{equation} 
The end result could be substitute back in \cref{eq:xi_eq_pphi} to get \cref{eq:xi_yi_eq_pphi_final}.

\subsection{Numerical Integration}
The terms $I_{x_{ij}},I_{y_{ij}}$ numerical evaluation will be presented below. 
\begin{multline}
	I_{x_{ij}}=\frac{\Gamma_i}{2\pi}\biggr[\int_{x_j+a}^{x_j-a}\frac{-(y_j-\sqrt{a^2-(x-x_j)^2}-y_i)}{(x-x_i)^2+(y_j-\sqrt{a^2-(x-x_j)^2}-y_i)^2}dx\\ + \int_{x_j-a}^{x_j+a}\frac{-(y_j+\sqrt{a^2-(x-x_j)^2}-y_i)}{(x-x_i)^2+(y_j+\sqrt{a^2-(x-x_j)^2}-y_i)^2}dx \biggr]
\end{multline}

\begin{multline}
	I_{y_{ij}}=\frac{\Gamma_i}{2\pi}\biggr[\int_{y_j+a}^{y_j-a}\frac{x_j+\sqrt{a^2-(y-y_j)^2}-x_i}{(y-y_i)^2+(x_j+\sqrt{a^2-(y-y_j)^2}-x_i)^2}dx\\ + \int_{y_j-a}^{y_j+a}\frac{x_j-\sqrt{a^2-(y-y_j)^2}-x_i}{(y-y_i)^2+(x_j-\sqrt{a^2-(y-y_j)^2}-x_i)^2}dx \biggr]
\end{multline}

\section{Comparison with Ragazzo's Hamiltonian}\label{app:Hamiltonian_Ragazzo}
It may be interesting to derive a Hamiltonian formulation from the Lagrangian one described in \cref{sec:Variational Dynamics of Rankine Vortex Patches as an Application}, relying on the Legendre transformation \citep{goldstein2002classical} between Lagrangian and Hamiltonian functions. The transformation is represented as
\begin{equation}\label{eq:Hamiltonian_from_legendre_trans}
	H(\bm{q}, \bm{p}, t)=\sum_i{\dot{q}_i p_i}-L(\bm{q}, \bm{\dot{q}}, t),
\end{equation}
where $p_i$'s are the generalized momenta given by: 
\begin{equation}\label{eq:generalzied_momenta}
	{p}_i=\frac{\partial L\left(\bm{q}, \bm{\dot{q}}, t\right)}{\partial {\dot{q}}_i}.
\end{equation}
Recall the derived Lagrangian in \cref{eq:Lag_flow+Lag_core} and substitute it into \cref{eq:generalzied_momenta}, then, the generalized momenta will read as
\begin{equation}\label{eq:generalzied_momenta_with lagnragnian}
	{p}_{i}=\frac{\partial L_{\phi_{t}}}{\partial {\dot{q}}_{i}}+\sum m_{i} {\dot{q}}_{i},
\end{equation}
where the first term is determined from \cref{eq:pLphi_pxid_simplified} (whose detailed computation is given in \cref{app:pphi}). The final form for the first term in \cref{eq:generalzied_momenta_with lagnragnian} includes a numerical integration over the cores circumferences, as shown in \cref{eq:Ix_simplified}. To simplify the Legendre transformation  analysis, and to allow for comparison with \cite{ragazzodensevortex_electro_simmilarity} Hamiltonian, point vortices with dense mass will be considered instead of vortex patches. As a result, the integrals $I_{x_{ij}}$, $I_{y_{ij}}$ defined in (\ref{eq:numerical_int_naming}) vanish. By this and relying on \cref{eq:Ix_simplified}, \cref{eq:generalzied_momenta_with lagnragnian} yields
\begin{equation}\label{eq:my_momenta}
	{p}_i=\sum m_{i} {\dot{q}}_{i}-\Gamma_i \rho \left({q}_i \times \bm{e}_3\right)	,
\end{equation}
and the Hamiltonian will be given by
\begin{equation}\label{eq:my_H}
	\mathcal{H}=\sum \frac{1}{2 m_i}\left\|{p}_i+ \Gamma_i \rho \left({q}_i \times \bm{e}_3\right)\right\|^2-T.
\end{equation}
Interestingly, the previous results are in agreement with the Hamiltonian and generalized momenta of \cite{ragazzodensevortex_electro_simmilarity}\footnote{There is a sign difference because of the arbitrary sign definition for the stream function.} as presented below. 
\begin{align}\label{eq:Ragazzo H and P}
	\mathbf{p}_j=&\sum m_{j} \mathbf{\dot{q}}_{j}+\frac{\Gamma_j \rho}{2}\left(\mathbf{q}_j \times \mathbf{e}_3\right)\\
	\mathcal{H}=&\sum \frac{1}{2 m_j}\left\|\mathbf{p}_j-\frac{\Gamma_j \rho}{2}\left(\mathbf{q}_j \times \mathbf{e}_3\right)\right\|^2+(-\rho W)
\end{align}
Also, setting the core size to zero in our formulation (so the integrals $I_{x_{ij}}$, $I_{y_{ij}}$ vanish) while considering non-zero core mass (i.e., considering dense point vortices), the resulting equations of motion are exactly the same as those by \cite{ragazzodensevortex_electro_simmilarity} using analogy with electromagnetism. However, the present analysis is more general, relying on first principles in fluid mechanics and can be generalized to other scenarios (time-varying vortices and other potential forces).

\end{document}